\documentclass[12pt]{article}
\usepackage{graphicx}
\usepackage{amsmath}
\usepackage{amssymb}
\begin{document}
\title{Estimation of eikonal parameter $c$ from the elastic scattering data.}
\author{V.A.Abramovsky, A.V.Dmitriev}

\maketitle
\begin{abstract}
In this paper we applicate eikonalization procedure to the processes of elastic scattering at high energies. We have shown, that elastic data is consistent with relatively low values of eikonal parameter $c_p < 1$  only. It is in contrast with theoretical considerations.
\end{abstract}

Let`s consider processes of elastic scattering in the framework of the eikonalization procedure. 

There is two view on the eikonalization procedure. At first, eikonalization was considered as a natural way to come from Regge behaviour $\sigma_{tot}(s) \sim s^{\Delta}$ to Froussoir-like behaviour $\sigma_{tot}(s) \sim ln(s)^2$. Both pure Regge model and black-disk model well describe total cross-section data. From the other hand, data on elastic cross-sections $\frac{d\sigma}{dt}(s,t)$ can be described assuming only one Regge pole (pomeron) at low and moderate $|t|<1GeV^2$. Here we will investigate possibility of eikonalization model to describe elastic scattering data at non-zero transferred momenta.

It is convenient to write up eikonalized amplitude as
\begin{equation}
\begin{array}{l}
A(s,\overline{k})=\frac{s}{i}\frac{4\pi}{c^2}\int \left\{e^{ic\chi(b)}-1\right\}J_0(bk)bdb\\
\chi(b)=\frac{1}{s}\int a(s,\overline{k})J_0(bk)bdb
\end{array}
\end{equation}
Here $\chi(b)$ is eikonal profile function and $a(s,\overline{k})$ is generic amplitude.

In the Regge model eikonal profile $\chi(b)$ can be calculated as
\begin{equation}
\chi(b)=-\frac{c^2}{8\pi}\frac{g_ag_b}{R_a^2+R_b^2+\alpha^{\prime}Y+i\pi\alpha^{\prime}/2}e^{-\frac{1}{4}\frac{b^2}{R_a^2+R_b^2+\alpha^{\prime}Y+i\pi\alpha^{\prime}/2}}e^{\Delta(Y+\frac{i\pi}{2})}
\label{chiR}
\end{equation}
To estimate eikonalization parameters, we will try to describe experimental data on total and elastic cross sections
\begin{equation}
\begin{array}{l}
\sigma_{tot}=\frac{1}{s}ImA(s,0)\\
\frac{d\sigma}{dt}=\frac{1}{16\pi s^2}|A(s,\overline{k})|^2\\
t=\overline{k}^2\\
\end{array}
\end{equation}
at the domain of $0.2GeV^2<t<1GeV^2$ and $\sqrt{s} \geq 53GeV^2$, there we have no non-vacuum reggeons and no perturbative $\frac{1}{t^{10}}$ behaviour.

At first stage we will start from conventional Regge picture, and will vary all parameters in (\ref{chiR}) at given $c$. We take only $\frac{d\sigma}{dt}$ data here to concentrate on the $t$ dependence of $A(s,t)$ and to avoid problems with phase of $A(s,t)$. As the result, we get function of likely-hood $\frac{\chi^2}{n.d.f.}$,  depending on $c$, see Fig.\ref{fig:chi_c}.
\begin{figure}
\begin{center}
\includegraphics[scale=0.5]{fig/chi.epsi}
\end{center}
\caption{Dependence of likely-hood function on c}
\label{fig:chi_c}
\end{figure}

\begin{figure}
\begin{center}
\includegraphics[scale=0.5]{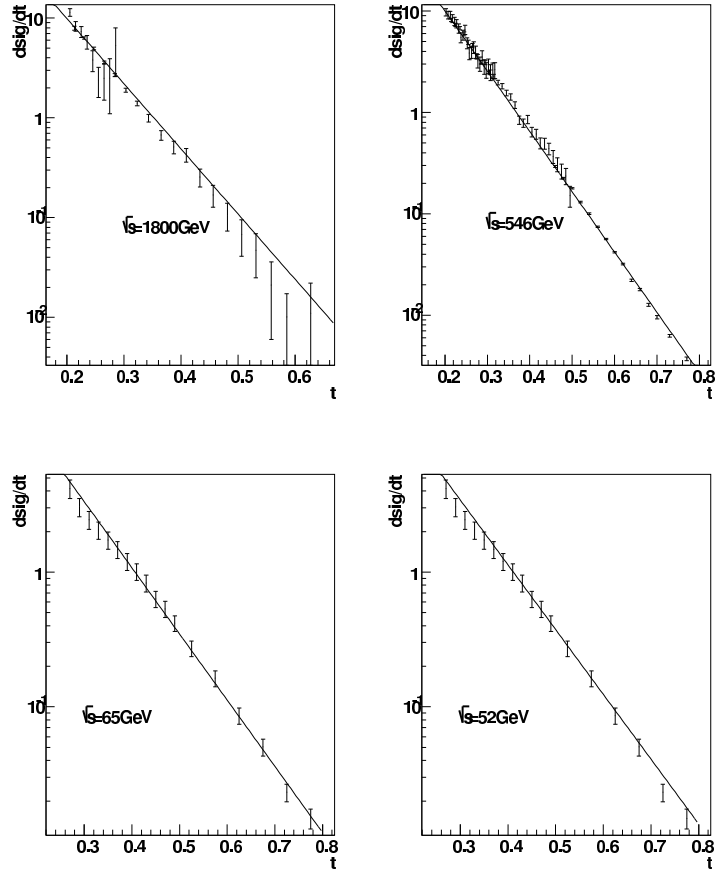}
\end{center}
\caption{$\frac{d\sigma}{dt}$ at $c=0$ (pure pomeron) and at optimal value $c=0.125$. Difference between graphics is not viewable, so we give one figure.}
\label{fig:el_0}
\end{figure}

\begin{figure}
\begin{center}
\includegraphics[scale=0.5]{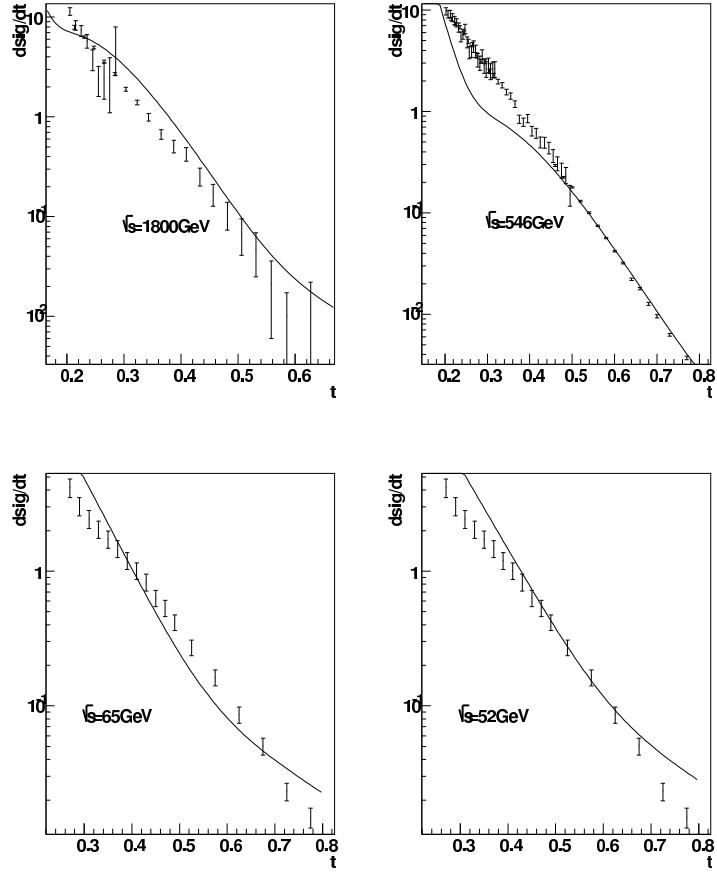}
\end{center}
\caption{$\frac{d\sigma}{dt}$ at $c=0.4$ (region of hardly divergence of experimental data and eikonal model)}
\label{fig:el_04}
\end{figure}

Dependence of $\frac{d\sigma}{dt}$ at some critical values of $c$ is given at Fig.\ref{fig:el_0} and Fig.\ref{fig:el_04}.

\newpage

It is clear, that this eikonal model with exponential pomeron vertex can describe experimental data only at low region $c \lesssim 0.2$. Including of total cross-section does not change this situation, function $\frac{\chi^2}{n.d.f.}$ rises even more quickly and has no minimum at non-zero $c$. Theory predict, that $c$ is high, $c \sim 1.5$ and there is strong limit $c>1$, so we have serious problem. We can try to reject exponential form of pomeron vertex, but we have no other assumptions about functional form of the vertexes.

It is natural to extract eikonal profile function $\chi(b)$ from experimental data  and test it regge-like behaviour. To make it, we approximate observed amplitude $A(s,t)$ by regge-like form:
\begin{equation}
\begin{array}{l}
A(s,t)=A(s,0)e^{B(s)t}\\
A(s,0)=s\sigma_{tot}(s)(\rho+i)\\
B(s)=\frac{B_{real}(s)}{2}+i\pi\frac{\alpha^{\prime}}{2}\\
\alpha^{\prime} \sim 0.25
\label{param}
\end{array}
\end{equation}
there $B_{real}(s)$ is logarithmic slope of $\frac{d\sigma}{dt}$ at given $s$. Reliable of this parametrisation is clear from Fig.\ref{fig:el_04}. We have only to test stability of our results on the changes of unobservable imaginary part of $\beta(s)$.

Calculations is trivial enough, and we have for the generic amplitude
\begin{equation}
a(s,\overline{k})=\frac{4\pi s}{ic^2}\int ln \left(1+\frac{ic^2A(s,0)}{8\pi sB(s)}e^{-\frac{b^2}{4B}} \right)J_0(bk)bdb
\end{equation}
Imaginary part of logarithm we define to be small at high $b$ and to be continuous at all $b$.

\begin{figure}
\begin{center}
\includegraphics[scale=0.5]{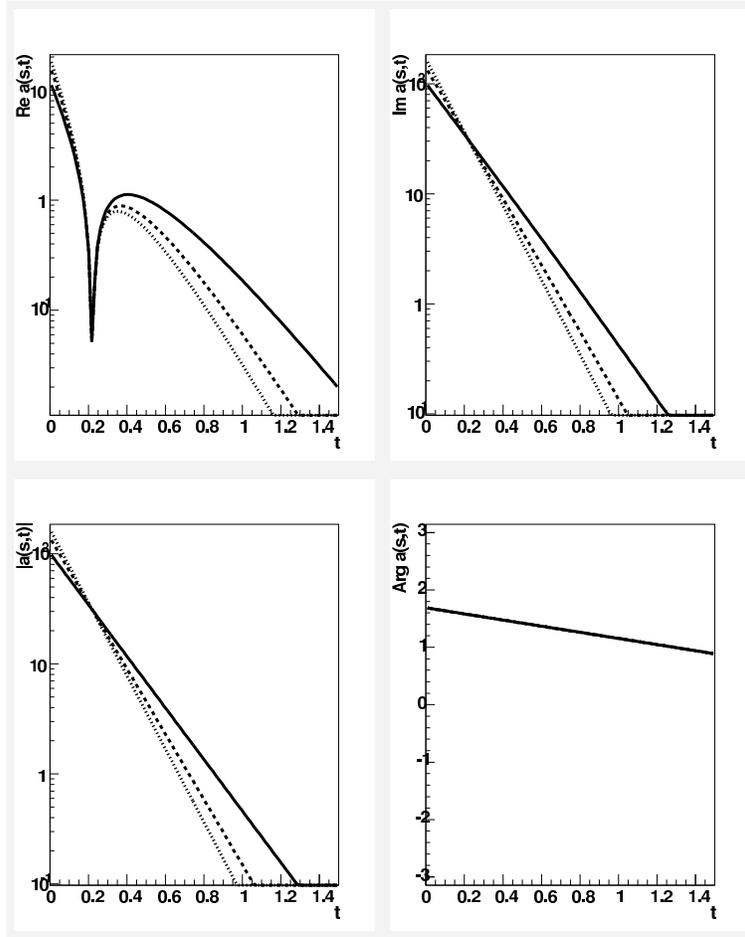}
\end{center}
\caption{Generic amplitude $a(s,t)$, divided on $s$, reconstructed for c=0 at energies $\sqrt{s}=1800 GeV$ (dotted line),  $\sqrt{s}=546 GeV$ (dashed line), $\sqrt{s}=65 GeV$ (solid line)}
\label{gener0}
\end{figure}

\begin{figure}
\begin{center}
\includegraphics[scale=0.5]{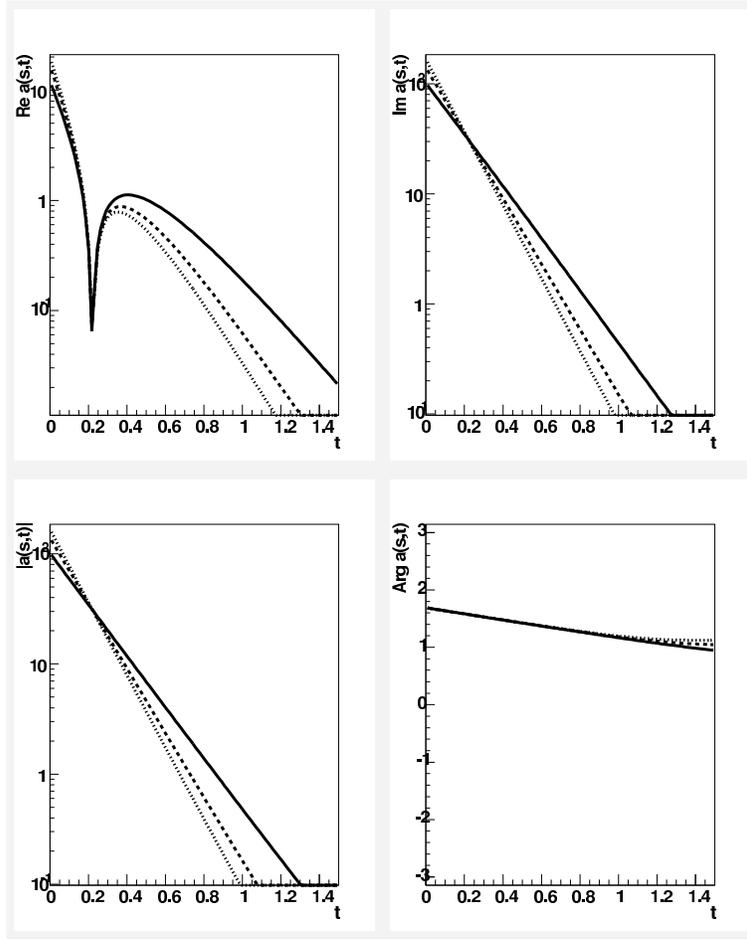}
\end{center}
\caption{Generic amplitude $a(s,t)$, divided on $s$, reconstructed for c=0.125 at energies  $\sqrt{s}=1800 GeV$ (dotted line),  $\sqrt{s}=546 GeV$ (dashed line), $\sqrt{s}=65 GeV$ (solid line)}
\label{gener0.125}
\end{figure}

Dependence of the reconstructed generic amplitude on $t$ is shown at Figs.\ref{gener0}-\ref{gener1.2}.

\begin{figure}
\begin{center}
\includegraphics[scale=0.5]{fig/gen0.75.epsi}
\end{center}
\caption{Generic amplitude $a(s,t)$, divideded on $s$, reconstructed for c=0.75 at energies  $\sqrt{s}=1800 GeV$ (dotted line),  $\sqrt{s}=546 GeV$ (dashed line), $\sqrt{s}=65 GeV$ (solide line)}
\label{gener0.75}
\end{figure}

\begin{figure}
\begin{center}
\includegraphics[scale=0.5]{fig/gen1.2.epsi}
\end{center}
\caption{Generic amplitude $a(s,t)$, divideded on $s$, reconstructed for c=0.75 at energies  $\sqrt{s}=1800 GeV$ (dotted line),  $\sqrt{s}=546 GeV$ (dashed line), $\sqrt{s}=65 GeV$ (solide line)}
\label{gener1.2}
\end{figure}

\newpage

\begin{figure}
\begin{center}
\includegraphics[scale=0.7]{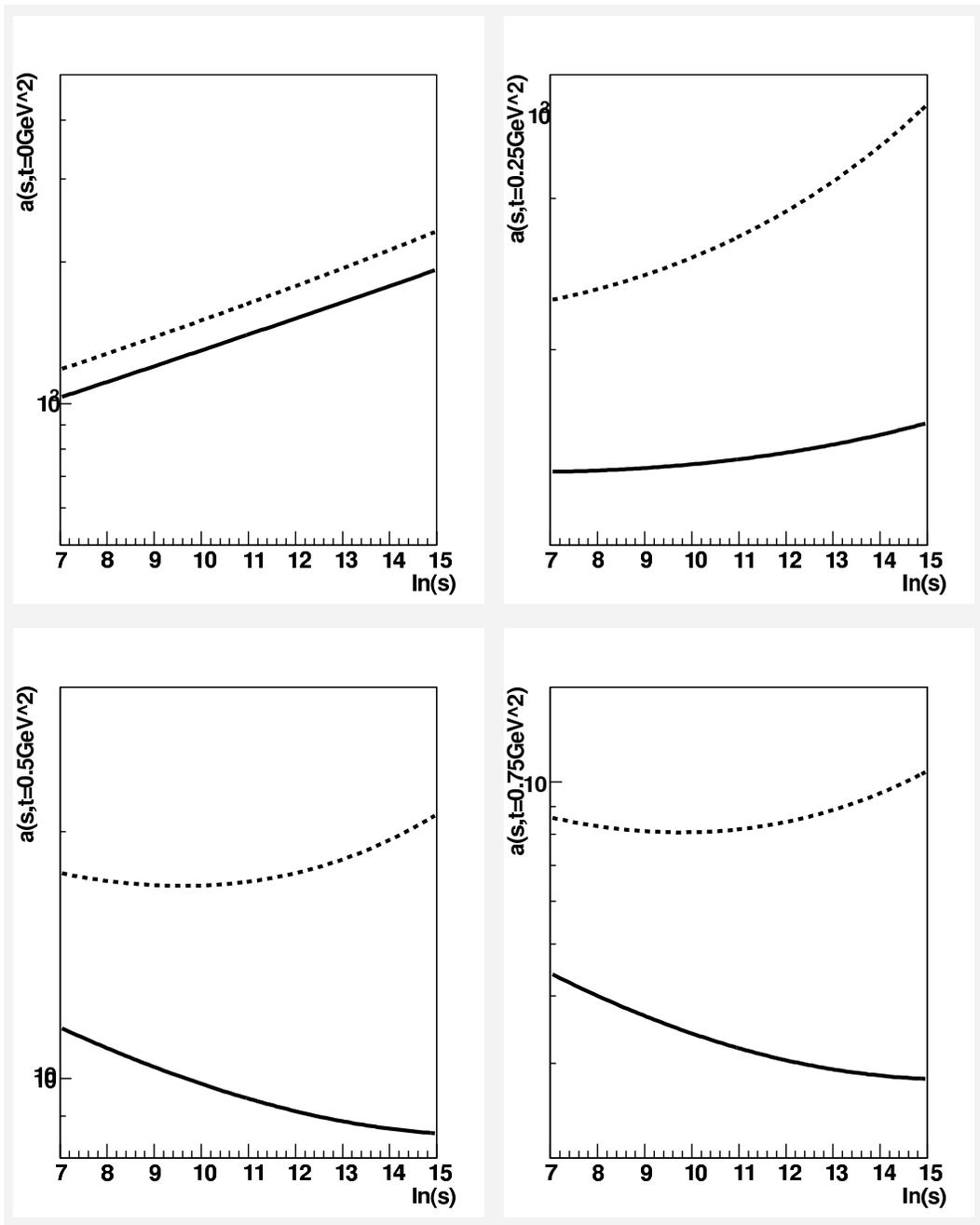}
\end{center}
\caption{Dependence of $a(s,t)$ on $ln(s)$ at $c=0.75$ (solid line) and $c=1$ (dashed line)}
\label{gen_s}
\end{figure}

From the canonical Regge theory, we expect for $a(s,t)$ to be
\begin{equation}
a(s,t)=e^{i\frac{\pi\alpha(t)}{2}}\beta(t)s^{\alpha(t)}
\label{Regge}
\end{equation}
there residue $\beta(t)$ is real.
We see, that our reconstructed amplitudes does not satisfy to equation (\ref{Regge}) at high $c$, because we have $s$-depended phases and (it is more serious), zeroes moving with $s$. Moreover, it is clear from analyse of $s$-dependence of $a(s,t)$ (see Fig.\ref{gen_s}), that $a(s,t)$ does not scale as  $s^{\alpha(t)}$ at high $c$.

\newpage

This results is stable under our assumptions about the phase of $A(s,t)$, at $c>0.75$ we can choose any small imaginary part of $B(s)$ in equations (\ref{param}) without any influence on the reconstructed amplitude $a(s,t)$. It is possible, that $s$-dependence phase of $a(s,t)$ is arisen from wrong assumptions about phase of the $A(s,t)$, but moving zeroes is stable at high $c$ ($c \geq 1$).

We can consider most general properties of the pomeron cuts. Many theoretical models (see \cite{collins} for review) leads us to the next form for the two-pomeron cut
\begin{equation}
\begin{array}{l}
A^c_2(s,t)=\frac{i}{16\pi^2s}\int^{0}_{-\infty}dt_1dt_2\frac{\theta(-\lambda(t,t_1,t_2))}{(-\lambda(t,t_1,t_2))^{1/2}}(N(t,t_1,t_2))^2 A^P(s,t_1)  A^P(s,t_2) \\
\lambda(t,t_1,t_2)=t^2+t_1^2+t_2^2-2(tt_1+tt_2+t_1t_2)
\end{array}
\end{equation}
Amplitude $A^P(s,t)$ is mostly imaginary, so cut $A^c_2$ have opposite sign to pole $A^P(s,t)$. So, at high enough $N(t,t_1,t_2)$ cuts will lead to changing of the exponential generic form of the pole (log-plot became non-linear with up-to convexity) and to existence of zeroes, which is moving with $s$. No of this phenomena is observed.

 So, we can state, that eikonalization parameter $c$ is limited by $c<0.5$, if we assume exponential form of the generic pomeron and by $c<1$ if we make no assumptions about form of the pomeron residue.  Anyway, this results is in contrast with the theory expectations and strongly limit eikonalization models, such as restoring of unitarity and renormalization of the pomeron flux in the single diffraction.

\section*{Acknowledgments}
We thank N.V.Prikhod`ko for useful discussions. This work was supported by RFBR Grant RFBR-03-02-16157a and Grant of Ministry for educations E02-3.1-282. Dmitriev A.V. was supported by S.-Petersburg grant for young scientists and specialists M04-2.4K-364.

\end{document}